# Modeling of Self-Pumped

# Singly Resonant Optical Parametric Oscillator

**Chengxian Deng**

**Physics Department, South China University of Technology (SCUT),**

**Guangzhou, 510641, P. R. China**

*Corresponding author: dengchx@scut.edu.cn*

**Abstract:**

A model of the steady-state operating, self-pumped singly resonant optical parametric oscillator (SPSRO) has been developed. The characteristics of quasi three-level laser gain medium pumped longitudinally have been taken into account. The characteristics of standing wave cavity, reabsorption losses, focusing Gaussian beams of the pump laser, fundamental laser and signal wave have been considered in the analyses. Furthermore, The power characteristics of threshold and efficiency have been analyzed, employing a $Yb^{3+}$-doped periodically poled lithium niobate co-doped with MgO ($Yb^{3+}$:MgO:PPLN) as the medium of laser gain and second-order nonlinear crystal.

Keywords: Self-Pumped Singly Resonant Optical Parametric Oscillator (SPSRO), Quasi Three-Level Laser, Reabsorption

## 1. Introduction

Singly resonant optical parametric oscillator (SRO) has the advantage of high stability, but the pump threshold is high. Employing intracavity optical parametric oscillator, one can make full use of the high intensity of the pump laser inside the cavity. There are many theoretical and experimental works about the intracavity SRO [1-3]. For these intracavity SROs, the laser gain medium in which the pump field of SRO is generated, and the second-order nonlinear crystal are two kinds of independent mediums. In recent years, the self-pumped singly optical parametric oscillators (SPSROs) attract people's interest [4-7]. One of their characteristics is that the pump field of the optical parametric oscillator is generated in the second-order nonlinear host crystal doped with laser-active ions. Compared with the conventional intracavity optical parametric oscillator, one of their merits is that one can build a compact tunable coherent radiation source with SPSRO. Combining with periodically poled host crystal, one can realize a low pump threshold and high efficiency optical parametric oscillator.

The quasi three-level $Yb^{3+}$-doped host crystals are receiving increasing interest since the development of strained InGaAs diode lasers emitting near 980 nm [5,7,8], where the $Yb^{3+}$ ion has a strong absorption peak. Laser action in the near infrared region has been demonstrated in a number of $Yb^{3+}$-doped materials [5, 7-9]. $Yb^{3+}$-doped solid state lasers show some advantages over $Nd^{3+}$-doped ones [8]: (1) The quantum efficiency of $Yb^{3+}$ is practically unity and the smaller quantum defect of $Yb^{3+}$ compared to $Nd^{3+}$, makes it suitable for high average power lasers. (2) The radiative lifetime of the laser level ranges from 0.4 to 2.3 ms, which implies greater energy storage efficiency. As a disadvantage, $Yb^{3+}$-doped lasers operate in a quasi three-level scheme which leads to reabsorption losses and temperature related effects. The SPSROs based on the second-order nonlinear host crystal doped with four-level and quasi three-level laser-active ions had been reported experimentally [4-7]. The typical four-level and quasi three-level laser-active ions are $Nd^{3+}$ and $Yb^{3+}$, respectively, and the typical host crystal is lithium niobate.

For the conventional intracavity optical parametric oscillator, the crystals length of laser gain medium and second-order nonlinear crystal can be independent tailored, a longer second-order nonlinear crystal is always required to reduce the pump threshold. But for a SPSRO, the second-order nonlinear crystal also acts as laser gain medium, the length cannot be independent tailored, a longer second-order nonlinear crystal means a longer laser gain medium, but leads to larger reabsorption losses in the quasi three-level laser gain medium. The crystal length and the losses must be balanced. For the collimated Gaussian beams and uniform pumped four-level laser gain medium, the results about threshold and power characteristics which presented in Refs [1] and [2] are approximately suitable for a SPSRO. But for the continuous wave or steady-state operating SPSROs pumped longitudinally, the strong focused beams are always required. The models developed in Refs [1] and [2] must be modified to suit for the operating conditions of strong focused beams, lasing in a quasi three-level gain medium that also acts as the second-order nonlinear crystal.

In this paper, to deal with the problem described in this section, we will follow closely with the results presented in Refs [10-13] to develop a model describing the steady-state operating, self-pumped singly resonant optical parametric oscillator (SPSRO). Our results are valid for small second-order nonlinear effects and no ground state depletion in laser gain medium. To consider ground state depletion, one must resort to the full numerical analyses [14]. In our analysis, the characteristics of longitudinally pumped quasi three-level laser gain medium, focused Gaussian beams of the pump laser, fundamental laser and signal wave, phase-mismatching have been considered in the derivations. Employing a $Yb^{3+}$-doped periodically poled lithium niobate co-doped with MgO ($Yb^{3+}$:MgO:PPLN), the characteristics of SPSROs based on quasi three-level laser gain medium will be investigated.

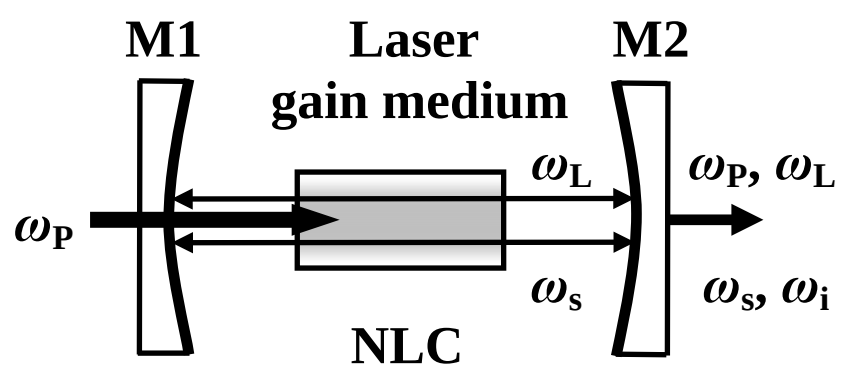

Fig. 1. Schematic diagram of a self-pumped singly resonant optical parametric oscillator, NLC: nonlinear crystal.

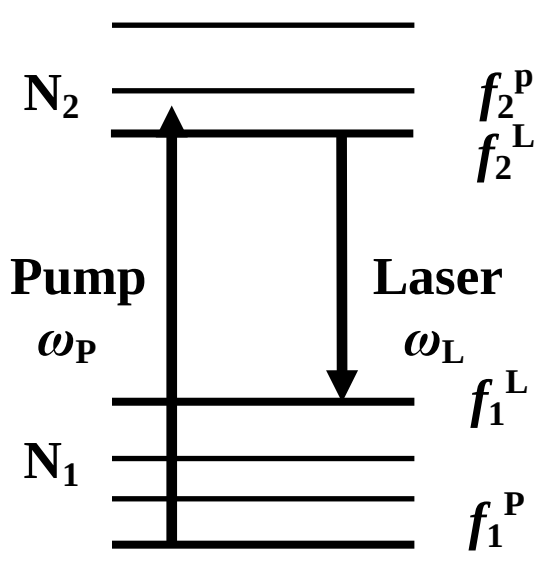

Fig. 2. Energy levels diagram of a quasi-three level laser gain medium.

## 2. Theory and Modeling

The scheme for longitudinally pumped SPSRO is shown in Fig. 1. For convenience, the external pump field for laser gain medium is called as external pump laser, the angular frequency is $\omega_P$. The laser generated in the laser gain medium is called as fundamental laser, the angular frequency is $\omega_L$. The fields generated in the second-order nonlinear crystal are called signal wave and idler wave, respectively, the respective angular frequency are $\omega_s$ and $\omega_i$, where $\omega_s+\omega_i=\omega_L$. For a SPSRO, the fundamental laser and signal field are resonating in the cavity.

To avoid the boundary problem, we assume that the mediums in which there are no lasing and second-order nonlinear interactions have the same refractive indices as the second-order nonlinear crystal has. The crystal length is $L_\chi$. To be more general, we assume that the cavity length of fundamental laser and the one of SRO are $L_{laser}$ and $L_{SRO}$, respectively. If the fundamental laser and SRO share the same cavity, $L_{laser}$ is equal to $L_{SRO}$. There are three kinds of threshold: (1) the external pump threshold for $Yb^{3+}$ lasing is called as fundamental laser threshold;

(2) the intracavity fundamental laser power reaches the conventional SRO threshold is called shortly as SRO threshold; (3) If the intracavity fundamental laser power reaches the conventional SRO threshold, the corresponding external pump power is called as SPSRO threshold. The subscripts letters *P*, *L*, *s*, *i* are denoted as external pump laser, fundamental laser, signal wave and idler wave, respectively.

The forward electric field of the fundamental laser in the standing-wave cavity is defined by

$$E_L^{\omega+}(x,y,z,t) = \frac{1}{2} A_L(t)\exp(-i\omega_L t) f_L^+(x,y,z)\exp(+ikz) + c.c. \tag{1}$$

where $A_L(t) = |A_L(t)|\exp(-i\varphi_L{}^+)$ is a single pass amplitude, other parameters are given in Appendix A. A similar definition is applied to the backward electric field. The similar definitions are also applied to the electric field of signal wave; expect that the subscript letter *L* is replaced by *s*. The nonlinear polarizations at the idler, fundamental laser and signal wavelength can be calculated using the methods presented in Refs [12] and [13]. They are respectively given by

$$P_{iNL}{}^+(x,y,z,t) = 2\varepsilon_0 d E_L{}^+(x,y,z,t) E_s{}^{+*}(x,y,z,t) \tag{2a}$$

$$P_{LNL}{}^+(x,y,z,t) = 2\varepsilon_0 d E_s{}^+(x,y,z,t) E_i{}^+(x,y,z,t) \tag{2b}$$

$$P_{sNL}{}^+(x,y,z,t) = 2\varepsilon_0 d E_L{}^+(x,y,z,t) E_i{}^{+*}(x,y,z,t) \tag{2c}$$

where *d* is the effective second-order nonlinear coefficient. $E_i{}^+(x,y,z,t)$ is the electric field of idler wave generated in the second-order nonlinear interaction in the forward direction of propagation.

Following the methods presented in Ref. [10], we can derivate the temporal evolution equation of the field of fundamental laser

$$\frac{d|A_L(t)|}{dt} + \frac{|A_L(t)|}{2\tau_{Lcavity}} = \left(-\frac{\omega_L}{2\varepsilon_0 n_L{}^2}\right)\mathrm{Im}\left\{\frac{\iiint\limits_{crystal}\left\{\begin{array}{l}[P_{laser}{}^+(x,y,z,t)+P_{LNL}{}^+(x,y,z,t)]\exp(i\varphi_L{}^+)h_L{}^+(x,y,z)\\ +[P_{laser}{}^-(x,y,z,t)+P_{LNL}{}^-(x,y,z,t)]\exp(i\varphi_L{}^-)h_L{}^-(x,y,z)\end{array}\right\}dV}{\iiint\limits_{cavity}|h_L(x,y,z)|^2\,dV}\right\} \tag{3}$$

where $P_{\text{laser}}^{+}(x,y,z,t)$ and $P_{\text{laser}}^{-}(x,y,z,t)$ are the polarizations generated in the lasing process at the forward and backward directions of propagation, respectively. For a conventional pump-resonant, signal-resonant optical parametric oscillator, we substitute the new parameters presented in Appendix A into the temporal evolution equations of the fields [1,10,12,13], and obtain the following equations

$$\frac{d\,|A_L(t)|}{dt}+\frac{|A_L(t)|}{2\tau_{Lcavity}}=\left(-\frac{\omega_L}{2\varepsilon_0 n_L^{\ 2}}\right)\left(\frac{4\varepsilon_0\pi d^2 L_\chi^{\ 2}}{L_{laser}\lambda_i n_i}\right)H_L\,|A_s(t)|^2|A_L(t)| \tag{4}$$

$$\frac{d\,|A_s(t)|}{dt}+\frac{|A_s(t)|}{2\tau_{scavity}}=\left(+\frac{\omega_s}{2\varepsilon_0 n_s^{\ 2}}\right)\left(\frac{4\varepsilon_0\pi d^2 L_\chi^{\ 2}}{L_{SRO}\lambda_i n_i}\right)\frac{H_L}{m_s^{\ 2}}|A_L(t)|^2|A_s(t)| \tag{5}$$

For a longitudinally pumped quasi three-level laser (only with laser action) [10, 11], the laser rate equations are well known, they can be rewritten as

$$\begin{aligned}\frac{dN_1(x,y,z,t)}{dt}=&-f_1^{\ L}R\frac{|f_P(x,y,z)|^2}{V_{P\alpha}}-\frac{N_1(x,y,z,t)-N_1^0(x,y,z)}{\tau}\\&+f_1^{\ L}c\sigma\frac{N_2(x,y,z,t)-N_1(x,y,z,t)}{n_L}\Phi_L(t)\frac{|f_L(x,y,z)|^2}{V_L}\end{aligned} \tag{6}$$

$$\begin{aligned}\frac{dN_2(x,y,z,t)}{dt}=&+f_2^{\ L}R\frac{|f_P(x,y,z)|^2}{V_{P\alpha}}-\frac{N_2(x,y,z,t)-N_2^0(x,y,z)}{\tau}\\&-f_2^{\ L}c\sigma\frac{N_2(x,y,z,t)-N_1(x,y,z,t)}{n_L}\Phi_L(t)\frac{|f_L(x,y,z)|^2}{V_L}\end{aligned} \tag{7}$$

$$\frac{d\Phi_L(t)}{dt}+\frac{\Phi_L(t)}{\tau_{Lcavity}}=\left(\frac{c\sigma}{n_L V_L}\right)\iiint_{crystal}\Delta N(x,y,z,t)\Phi_L(t)\,|f_L(x,y,z)|^2\,dV \tag{8}$$

where $N_1$ is the population density in the lower laser level, $N_2$ is the population density in the upper laser level, $\Delta N= N_2-N_1$ is the population inversion density, $N_1^{\ 0}$, $N_2^{\ 0}$ are the unpumped population densities for the two laser levels, $f_1^{\ \mathrm{L}}$ and $f_2^{\ \mathrm{L}}$ are the Boltzmann occupation factors for the lower and upper laser levels, respectively. Other parameters are written in Appendix A.

Associating with the rate equations of the quasi three-level laser and the equations of the fields for the optical parametric oscillator, using the dimensionless parameters defined in Appendix A, we obtain the rate equations of the fields of the SPSRO

$$\frac{d\Phi_L(t)}{dt}+\frac{\Phi_L(t)}{\tau_{Lcavity}}=\left(\frac{c\sigma_L}{n_L}\right)\frac{V_\chi}{V_L}\Phi_L(t)\cdot\iiint_{crystal}\Delta N(p,q,s,t)\,|\,f_L(p,q,s)|^2\,dpdqds-C_L H_L\Phi_s(t)\Phi_L(t) \tag{9}$$

$$\frac{d\Phi_s(t)}{dt}+\frac{\Phi_s(t)}{\tau_{scavity}}=C_L H_L\Phi_L(t)\Phi_s(t) \tag{10}$$

$$\begin{aligned}\frac{d\Delta N(x,y,z,t)}{dt}&=(f_1^L+f_2^L)R\frac{|\,f_P(p,q,s)|^2}{V_{P\alpha}}-\frac{\Delta N(p,q,s,t)-\Delta N^0(p,q,s)}{\tau}\\&\quad-(f_1^L+f_2^L)\frac{c\sigma_L\Delta N(p,q,s,t)}{n_L}\frac{|\,f_L(p,q,s)|^2}{V_L}\Phi_L(t)\end{aligned} \tag{11}$$

If a SPSRO is operating under the condition of continuous-wave or steady-state, and the power of the external pump laser is above the SPSRO threshold, the total number of fundamental laser photons $\Phi_L$ (two passes) in the cavity is constant $\Phi_{LSROth}$

$$\Phi_{LSROth}=\frac{1}{C_L H_L\tau_s} \tag{12}$$

The corresponding single pass power of the fundamental laser is

$$P_{LSROth}=\frac{T_s+\beta_s}{C_{LSRO}(\xi_L/m_s^{\,2})H_L} \tag{13}$$

From the equations (9),(10) and (11), we derive the power of the external pump laser $P_P$

$$P_P=(T_L+\beta_L)\frac{P_c}{\eta_P\eta_\alpha}\frac{1+B\left(\frac{L_{laser}}{L_\chi}\right)\left(\frac{V_\chi}{V_L}\right)\iiint_{crystal}\frac{|\,f_L(p,q,s)|^2}{1+Q_{LSROth}\,|\,f_L(p,q,s)|^2}dpdqds+Q_s\left(\frac{\tau_L}{\tau_s}\right)}{\left(\frac{V_\chi}{V_L}\right)\iiint_{crystal}\frac{|\,f_P(p,q,s)|^2|\,f_L(p,q,s)|^2}{1+Q_{LSROth}\,|\,f_L(p,q,s)|^2}dpdqds} \tag{14}$$

where $\eta_P$ is the quantum efficiency of pump process. $B$ is the ratio of reabsorption loss to fixed cavity loss of the fundamental laser. The dimensionless fundament laser photons and signal field photons are presented in Appendix A.

Below the SPSRO threshold, the power of the pump laser $P_P$ associated with the dimensionless power of the fundamental laser $Q_L$ (single pass) in the cavity is derived from the following equation

$$P_P = (T_L + \beta_L)\frac{P_c}{\eta_P \eta_\alpha} \frac{1 + B\left(\frac{L_{laser}}{L_\chi}\right)\left(\frac{V_\chi}{V_L}\right)\iiint_{crystal} \frac{|f_L(p,q,s)|^2}{1+Q_L\,|f_L(p,q,s)|^2} dpdqds}{\left(\frac{V_\chi}{V_L}\right)\iiint_{crystal} \frac{|f_P(p,q,s)|^2|f_L(p,q,s)|^2}{1+Q_L\,|f_L(p,q,s)|^2} dpdqds} \tag{15}$$

If $Q_L$ is equal to $Q_{LSROth}$, we can obtain the power of the SPSRO threshold $P_{PSPSROth}$.

The power threshold of the fundamental laser is derived from equation (15), and is given by

$$P_{Plaserth} = (T_L + \beta_L)\frac{P_c}{\eta_P \eta_\alpha} \frac{1 + B\left(\frac{L_{laser}}{L_\chi}\right)\left(\frac{V_\chi}{V_L}\right)\iiint_{crystal} |f_L(p,q,s)|^2\, dpdqds}{\left(\frac{V_\chi}{V_L}\right)\iiint_{crystal} |f_P(p,q,s)|^2|f_L(p,q,s)|^2\, dpdqds} \tag{16}$$

From the equation (14), we can derived the output power of signal wave $P_s$ (single pass), and it is written as

$$P_{sout} = T_s P_s = \eta_s (P_P - P_{PthSPSRO}) \tag{17}$$

the efficiency of output signal wave $\eta_s$ is

$$\eta_s = P_{LthSRO}\left(\frac{T_s}{\beta_s + T_s}\right)\left(\frac{\omega_s}{\omega_L}\right)\left(\frac{\eta_P\eta_\alpha}{P_C}\right)\left(\frac{V_\chi}{V_L}\right)\iiint_{crystal}\frac{|f_P(p,q,s)|^2|f_L(p,q,s)|^2}{1+Q_{LthSRO}\,|f_L(p,q,s)|^2} dpdqds \tag{18}$$

The total output power of idler wave generated in the two directions of propagation is

$$P_{iout} = C_{iSRO}\xi_S H_i P_S P_{LthSRO} \tag{19}$$

The output down-converted power is the sum of $P_{sout}$ and $P_{iout}$

$$P_{DCout} = P_{sout} + P_{iout} = \eta_s \left( P_P - P_{PthSPSROth} \right) \left( 1 + \frac{C_{iSRO} \xi_s H_i P_{LthSRO}}{T_s} \right) \tag{20}$$

We assume that for the SPSRO pumped by a laser with power value of $P_P$, the down-converted power is $P_{DCout}$. For this SPSRO structure, in the absence of singly resonant parametric process, under the same pump power, the corresponding output power of the fundamental laser is $P_{Lout}$. The ratio of $P_{DCout}$ to $P_{Lout}$ is

$$\eta_{DCout} = \frac{P_{DCout}}{P_{Lout}} = \frac{P_{sout} + P_{iout}}{P_{Lout}} \tag{21}$$

## 3. Numerical Results and Discussion

In the numerical calculation, we will employ the $Yb^{3+}$-doped periodically poled lithium niobate co-doped with MgO ($Yb^{3+}$:MgO:PPLN) as the laser gain medium and second-order nonlinear crystal [5], [7], [8], [15]. Without the further statements, we use the parameters of SPSRO presented in this paragraph. The impurity concentrations in the PPLN crystal are 1.2 at.% of ytterbium and 5mol% of MgO ($Yb^{3+}$ (1.2at.%):MgO(5mol%):PPLN). The operating temperature is 20˚C. The cw Ti:sapphire external pump laser at 980 nm is σ polarized (correspond to transitions from the lowest-lying Stark level of the $^2F_{7/2}$ state ($^2F_{7/2}(0)$) to the Stark level of the $^2F_{5/2}$ excited state ($^2F_{5/2}(0)$)). The fundamental laser at 1063 nm is π polarized (correspond to transitions from Stark level of the $^2F_{5/2}(0)$ excited state to the Stark level of the $^2F_{7/2}(3)$ state). The signal wave at 1360 nm is π polarized, and the corresponding π polarized idler wavelength is 4867.6nm; The refractive indices at the external pump laser wavelength, fundamental laser wavelength, signal wavelength and idler wavelength are 2.23, 2.15, 2.14 and 2.0, respectively. The $Yb^{3+}$ concentration is about $2.2\times10^{20}$ $cm^{-3}$. The value of the emission lifetime τ is 540μs. The pump saturation intensities for the ordinary and extraordinary polarized fields are 33.2 $kW/cm^2$ and 33.29 $kW/cm^2$, respectively, the corresponding absorption cross sections are $1.1314\times10^{-20}cm^2$ and $1.1284\times10^{-20}cm^2$, respectively. The values of the emission cross-section at fundamental laser wavelength are about $5.57\times10^{-21}cm^2$ for π polarization and $2.42\times10^{-21}$cm$^2$ for σ polarization. The Boltzmann occupation factors $f_1^{L}$ and $f_2^{L}$ are 0.0164 and 0.7522, respectively. The unpumped population inversion density $\Delta N^0=N_2^0-N_1^0$ is $3.62\times10^{18}$ cm$^{-3}$. The effective second-order nonlinear coefficient $d$ is 14 pm/V. The round trip losses of the fundamental laser and signal wave are $\beta_L=\beta_s=0.005$. The absorption coefficient of idler wave in the crystal is 1.05 cm$^{-1}$. The quantum efficiency of pump process $\eta_P$ is 1.0. The $M^2$ value of pump laser beam is 1.2. The cavity length is 10cm, and $L_{laser}$ is equal to $L_{SRO}$. The active crystal $L_\chi$ is 6mm. The positions of beams waist are in the centre of crystal. The radius of beam waist of fundamental laser $W_{0L}$ is 20 μm, $\xi_P$ and $\xi_s$ are equal to $\xi_L$, $R_L$ is 0.999, $R_s$ is 0.99, the output reflectivity of pump laser $R_P$ is 0.9.

Using the parameter values presented in this section, we will compare the results of our modeling with those presented in Ref. [5]. In the experiments of Ref. [5], the cavity length is about 100 mm, the pump and infrared fundamental beam waists were both of order 50 μm inside the crystal. From the experimental results, the power of the SPSRO threshold $P_{PSPSROth}$ was about 1.1 W. When the pump power incident on the cavity was 1.3 W, the laser was operated with an infrared output power of 30 mW at 1063 nm. The down converted output power was estimated to be 0.4 mW. From our calculation, the power of the SPSRO threshold $P_{PSPSROth}$ is 1.5 W. When the pump power incident on the cavity is 1.8W ($1.2P_{PSPSROth}$), the laser is operated with an infrared output power of 210 mW at 1063 nm, The down converted output power of signal and idler waves are 70 mW and 19 mW, respectively. Because the calculated results are sensitive to the parameter values, and these values did not give in detail in Ref. [5], furthermore, many factors of measurement were not taken into account, our results will not agree well with those experimental results.

(1) characteristics of different phase-mismatching

Figures 3(a), 3(b) and 3(c) show the variation of SPSRO threshold, efficiency of output signal wave and SRO threshold as functions of phase-mismatching parameters $\Delta kL_\chi$ with different crystal length $L_\chi$. As in a conventional SRO, for larger focusing parameter $\xi_L = L/(k_L W_{0L}^2)$, the SPSRO threshold is smaller for negative values of $\Delta kL_\chi$, and there is a minimum threshold for certain values of $\Delta kL_\chi$. In the following sections ((2)-(8)), the value of $\Delta kL_\chi$ at which SPSRO threshold is minimized is denoted by $(\Delta kL_\chi)_{min}$, and the corresponding SPSRO threshold $P_{PthSPSRO}$, efficiency of output signal wave $\eta_s$, SRO threshold $P_{LthSRO}$, output

power of signal wave $P_{\mathrm{sout}}$, down-converted power $P_{\mathrm{DCout}}$, down-converted efficiency $\eta_{\mathrm{DC}}$ and fundamental laser threshold $P_{\mathrm{Plaserth}}$ are plotted in the figures. For the parameters given in this section, the larger SPSRO threshold is, the higher efficiency of the output signal wave is. In general, the minimum value of SPSRO threshold and the maximum value of efficiency occur at different values of $\Delta kL_{\chi}$. In this paper, we will not perform the optimal analysis.

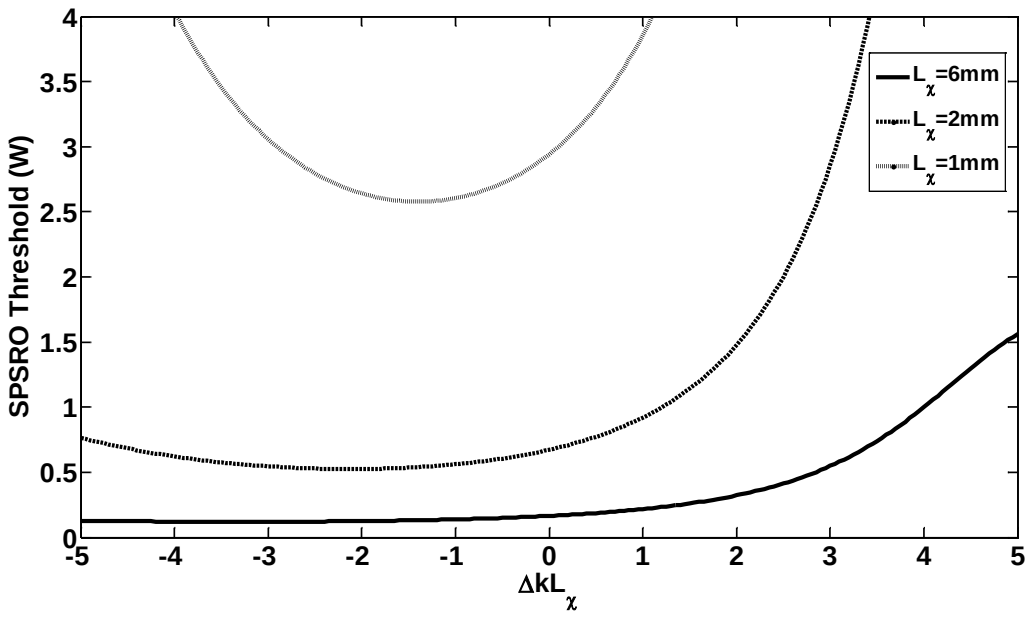

(a)

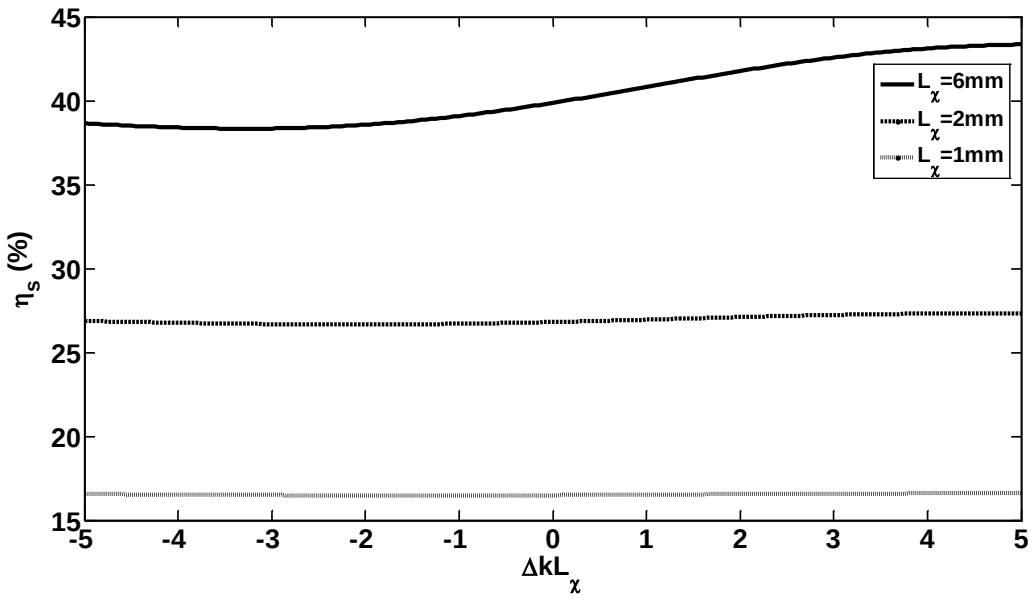

(b)

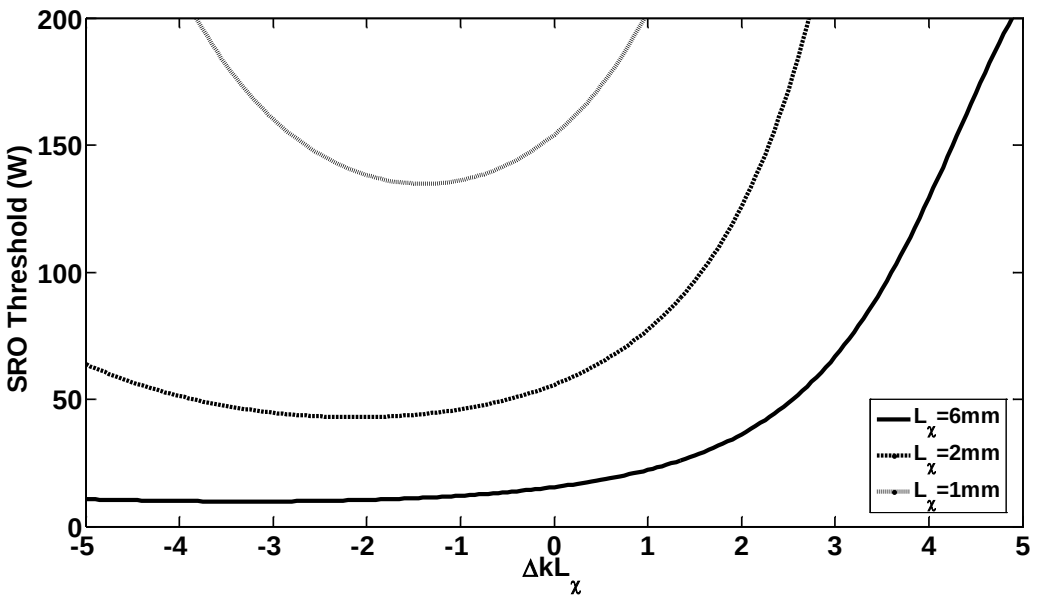

(c)

Fig. 3. (a), (b) and (c): Variation of SPSRO threshold, $\eta_s$ and SRO threshold as functions of $\Delta kL_{\chi}$ with different $L_{\chi}$.

(2) characteristics of different $Yb^{3+}$ concentration crystal

Figures 4(a), 4(b) and 4(c) show the variation of SPSRO threshold, efficiency of output signal wave and fundamental laser threshold as functions of crystal length $L_{\chi}$ with different $Yb^{3+}$ concentration crystal. The $Yb^{3+}$ concentration is expressed as the absorption coefficient of the ordinary polarization pump laser, and is given by $N_{\mathrm{Yb}}=\alpha_{\mathrm{PO}}/\sigma_{\mathrm{PO}}$. For $\alpha_{\mathrm{PO}}$=1.0, 2.0, 5.0, 10.0 /cm, the corresponding $Yb^{3+}$ concentration are 0.53, 1.33, 2.7, 5.33 at.%, respectively. The values of crystal length $L_{\chi}$ corresponding the lower threshold of SPSRO is about 6~8mm in the four kinds of $Yb^{3+}$ concentration, while the values of $L_{\chi}$ corresponding the lower threshold of conventional

laser is about 1mm. For a conventional SRO, the threshold is high, in general, is several Watt, thus one must employ a longer nonlinear crystal (cm order of magnitude) to reduce the SRO threshold. But for a quasi three-level laser, longer crystal leads to larger reabsorption loss, and the threshold is also increased. One must balance the crystal length and the $Yb^{3+}$ concentration. Similarly, for a SPSRO, the SRO threshold and net laser gain of fundamental laser must be balanced for an optimal consideration.

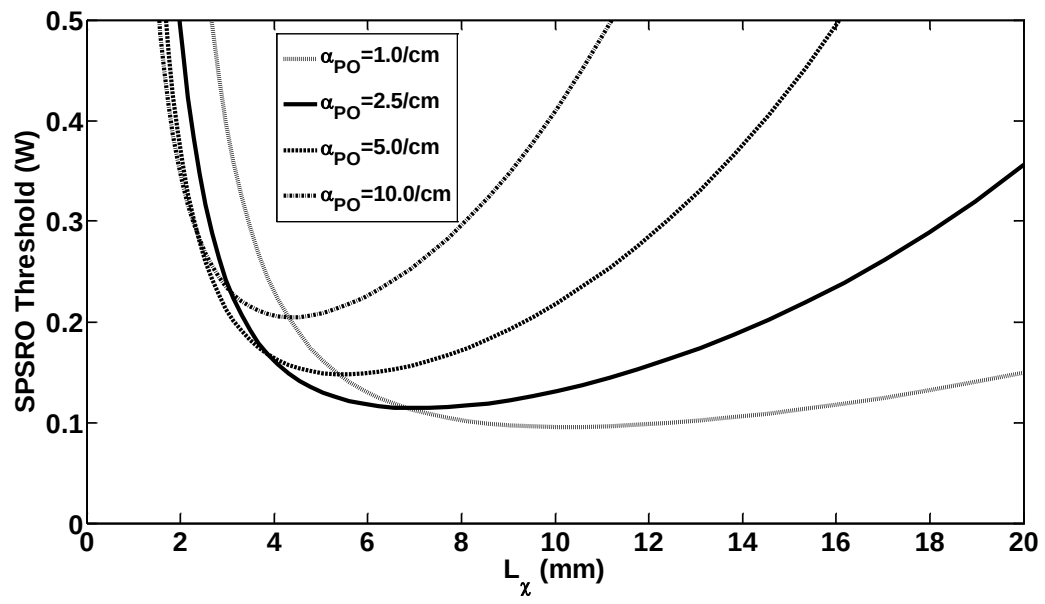

(a)

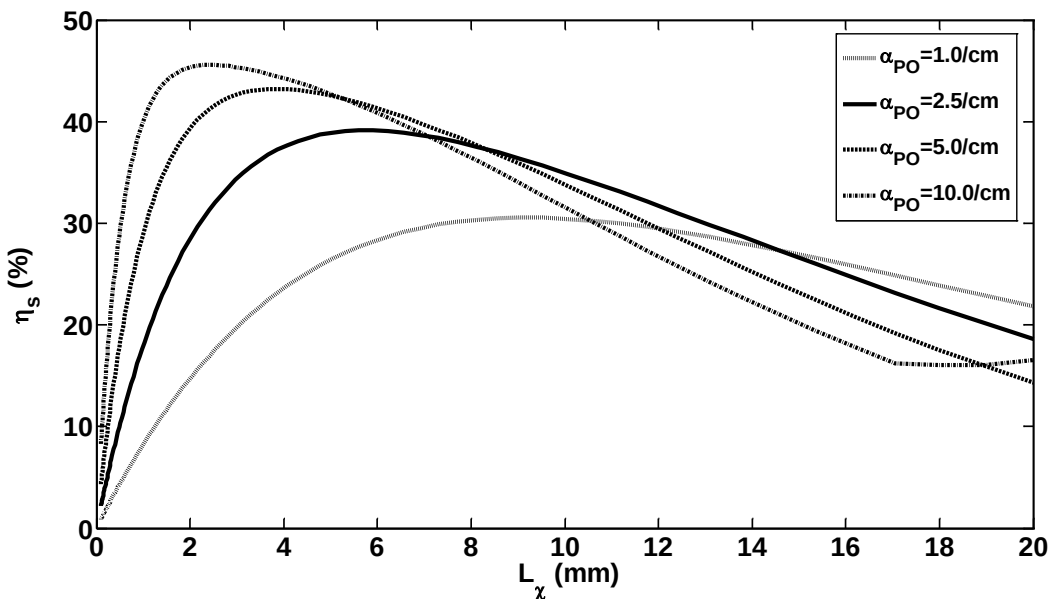

(b)

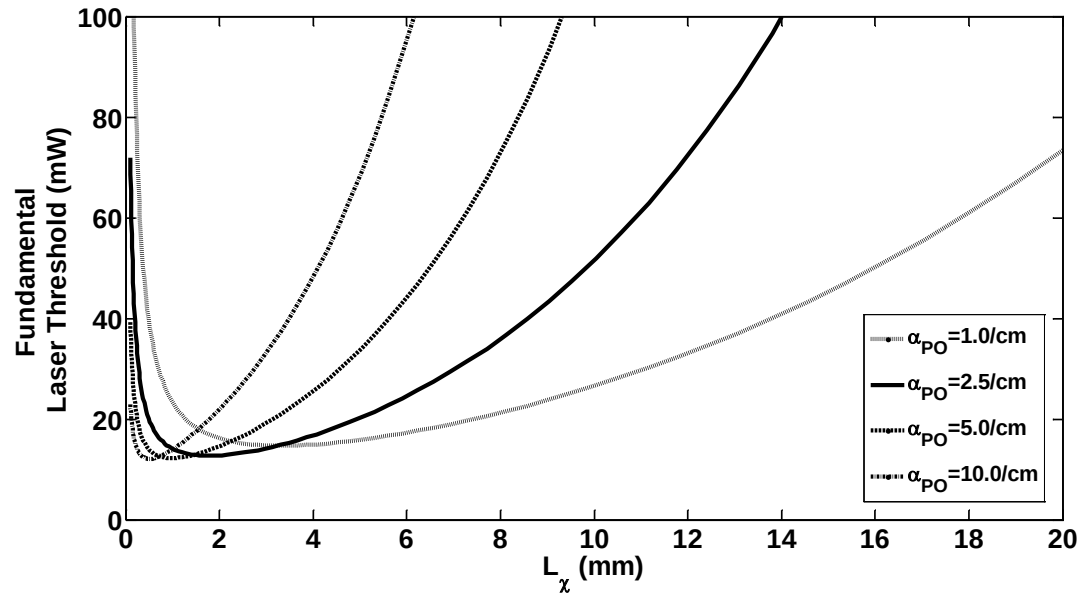

(c)

Fig. 4. (a), (b) and (c): Variation of SPSRO threshold, $\eta_s$ and fundamental laser threshold as functions of $L_\chi$ with different $Yb^{3+}$ concentration crystal.

(3) characteristics of different crystal temperature

Figures 5(a) and 5(b) show the variation of the SPSRO threshold and efficiency of output signal wave as functions of crystal length $L_\chi$ with different crystal temperature. Due to the properties of quasi three-level laser gain medium, higher crystal temperature leads to the effects of larger thermal populations of the ground and excited sub-levels, and leads to larger reabsorption losses, thus the SPSRO threshold is higher. But the efficiency of output signal wave does not change significantly for different crystal temperature.

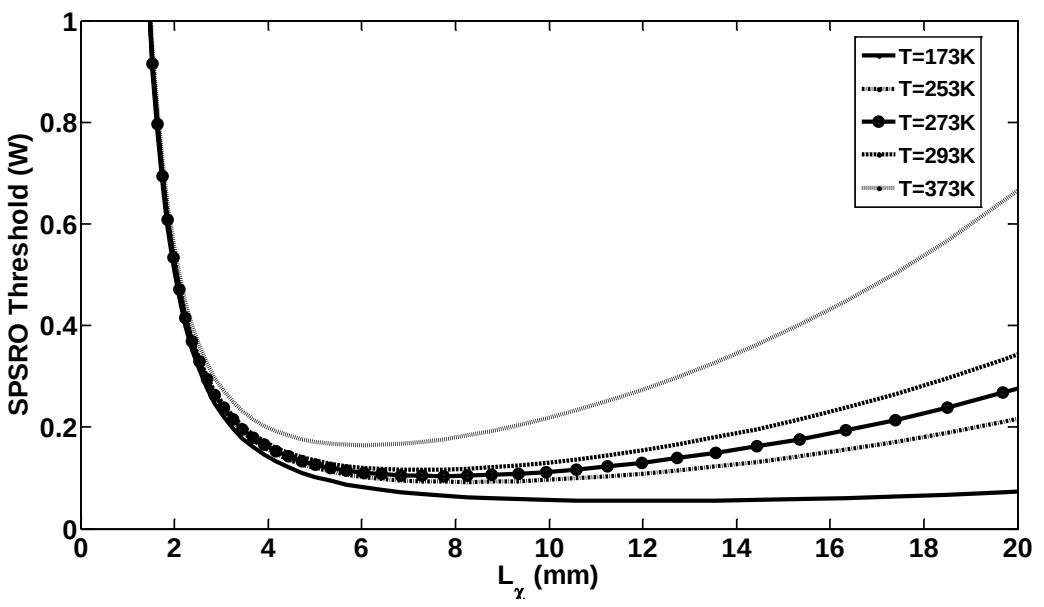

(a)

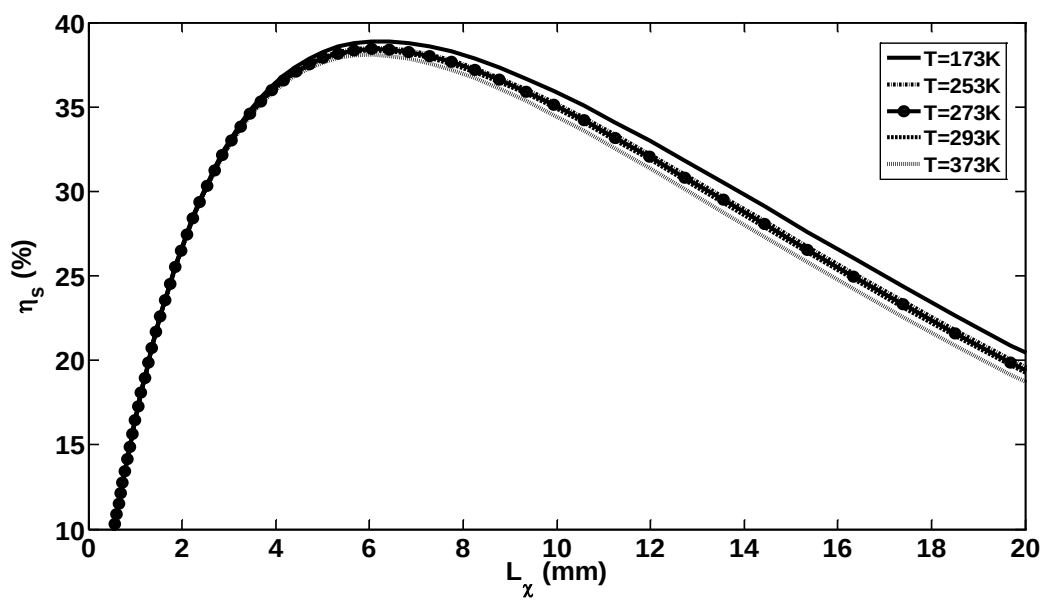

(b)

Fig. 5. (a) and (b): Variation of SPSRO threshold and $\eta_s$ as functions of $L_\chi$ with different crystal temperature.

(4) characteristics of different radius of beam waist of the fundamental laser $W_{0L}$

Figures 6(a), 6(b), 6(c) and 6(d) show the variation of SPSRO threshold, efficiency of output signal wave $\eta_s$, SRO threshold and fundamental laser threshold as functions of crystal length $L_\chi$ with different $W_{0L}$, the crystal temperature is 293K. While in the cases of figures 6(e), 6(f) and 6(g), the crystal temperature is 173K. As is expected, the crystal length at which the SPSRO threshold is minimum dose not consist with the one at which the fundamental laser threshold is minimum. Due to the diffractive effect and reabsorption loss, for smaller $W_{0L}$, the lower threshold is for smaller crystal length. The efficiency of output signal wave is smallest for $W_{0L}$=10μm in the four kinds of fundamental laser beam waist. As stated in the previous section, the efficiency of output signal wave does not change significantly for different crystal temperature, but the SPSRO threshold does.

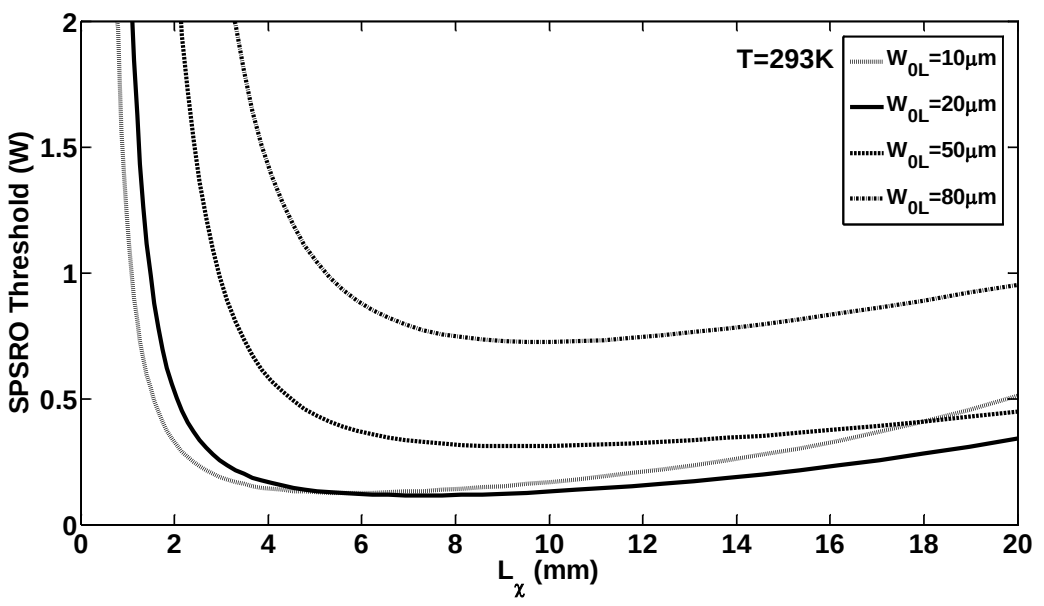

(a)

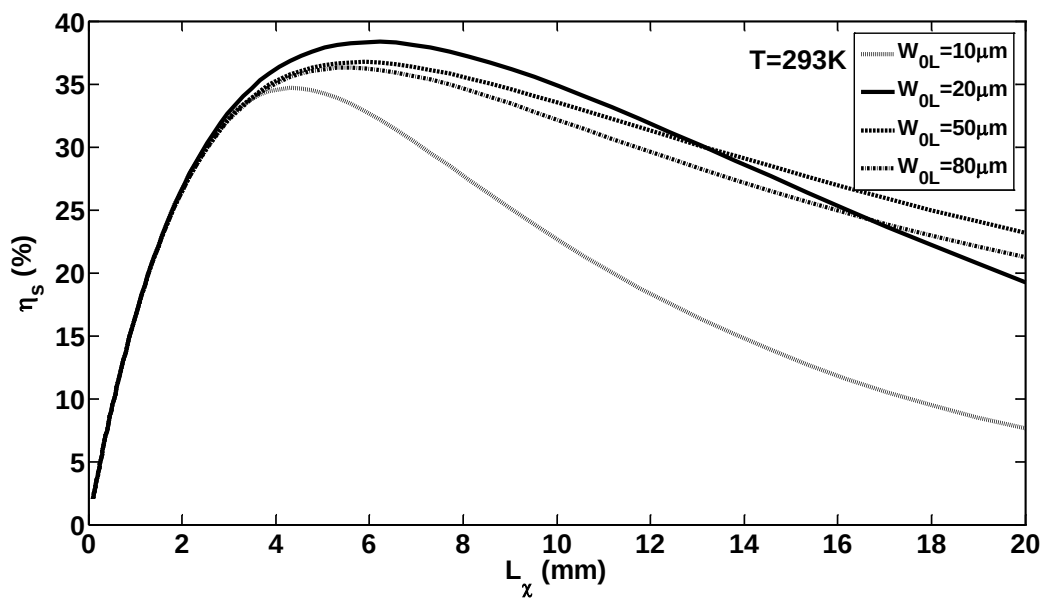

(b)

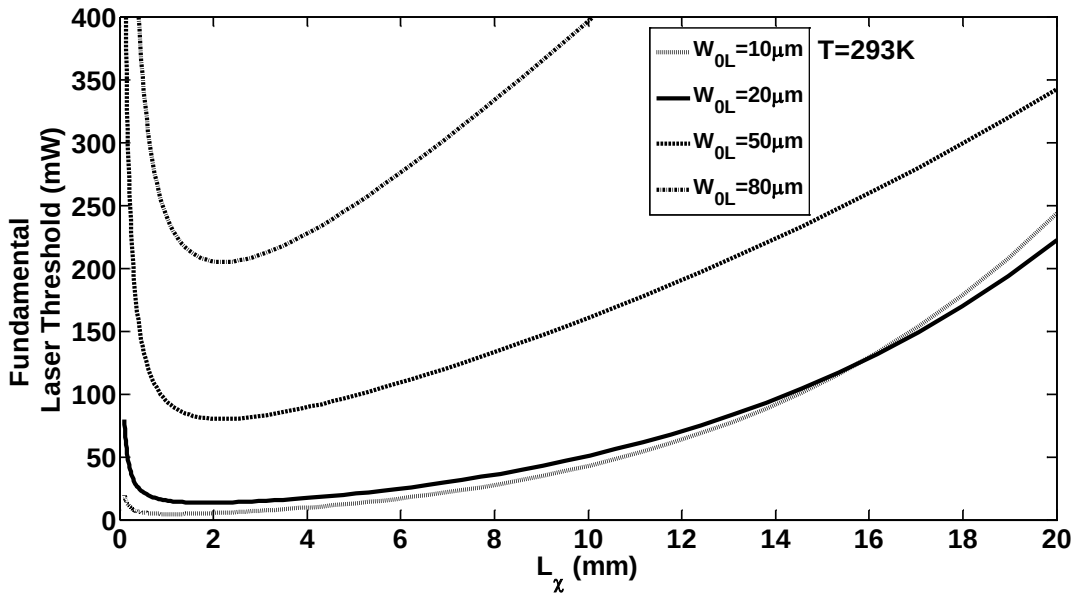

(c)

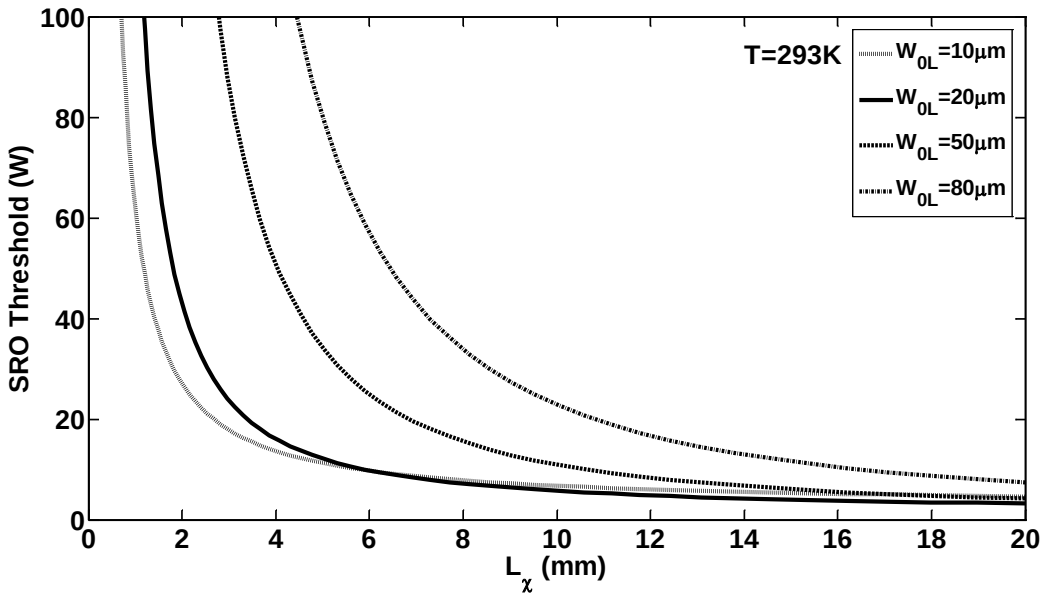

(d)

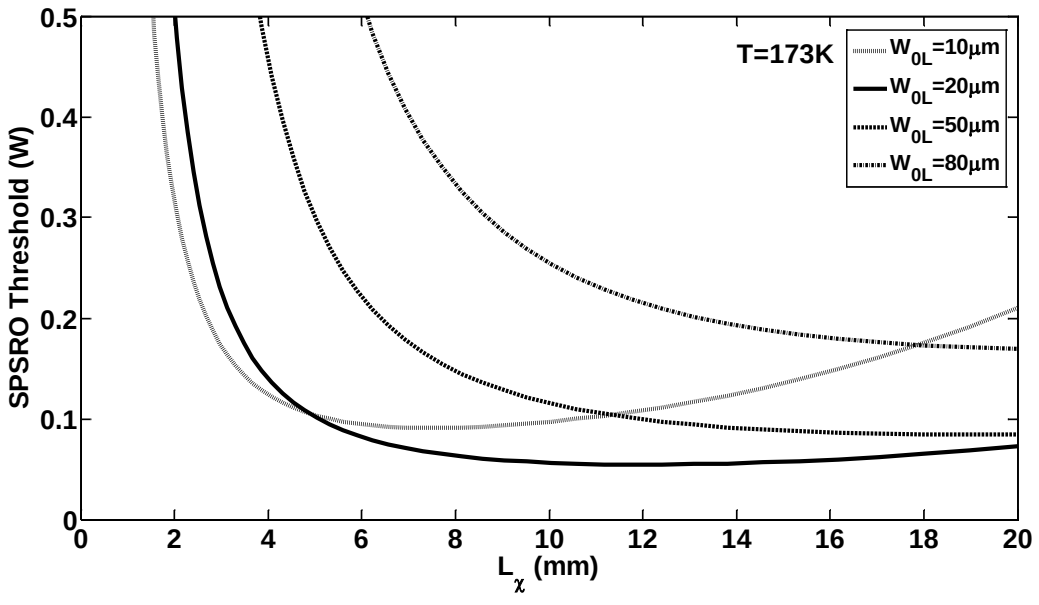

(e)

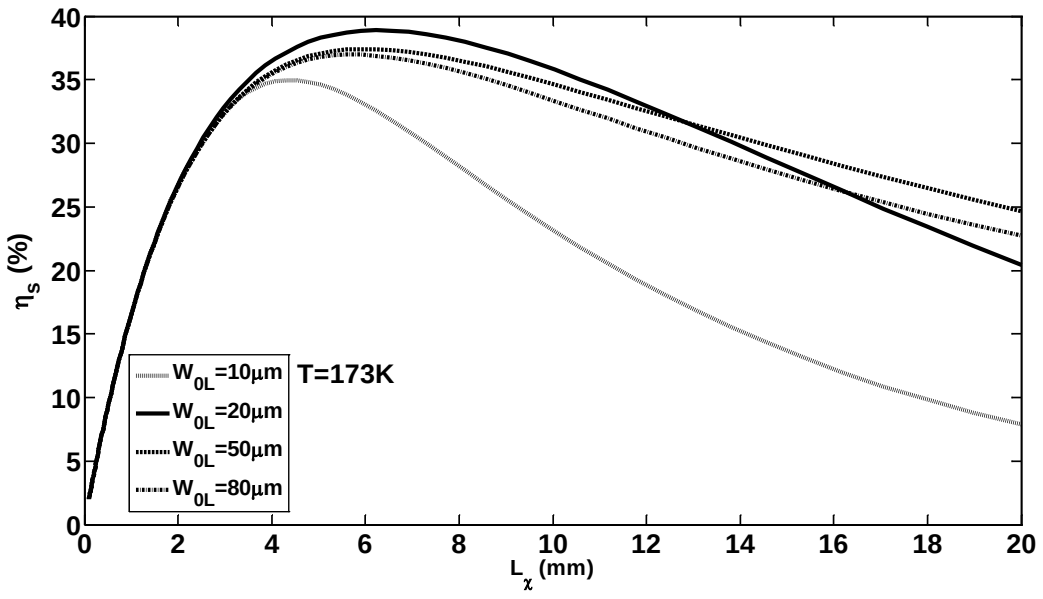

(f)

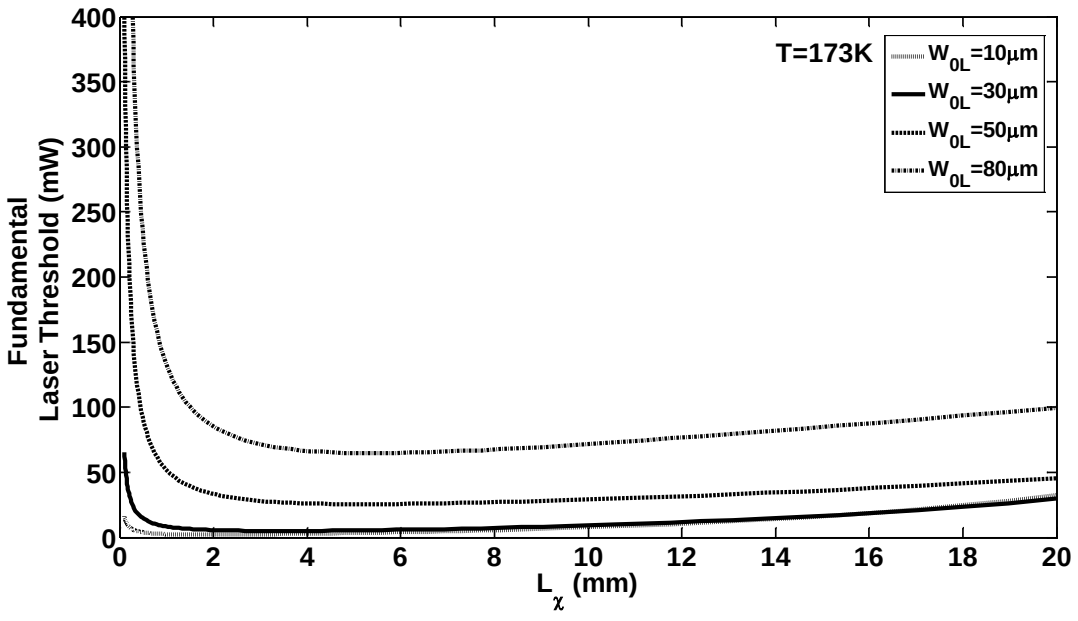

(g)

Fig. 6. Variation of SPSRO threshold, $\eta_s$, fundamental laser threshold and SRO threshold as functions of $L_\chi$ with different $W_{0L}$. (a), (b), (c), (d): crystal temperature is 293K. (e), (f), (g): crystal temperature is 173K.

(5) characteristics of different radius ratio of beam waist of the external pump laser to fundamental laser $m_P$

Figure 7(a) shows the variation of the output power of signal wave $P_{sout}$ as functions of the power of external pump laser $P_P$ with different $m_P$. For clarity, figure 7(b) shows the results that $P_P$ varies from 0.1 to 0.2W. From the figures, one can find that there is an optimal $m_P$ at which the SPSRO threshold is minimum.

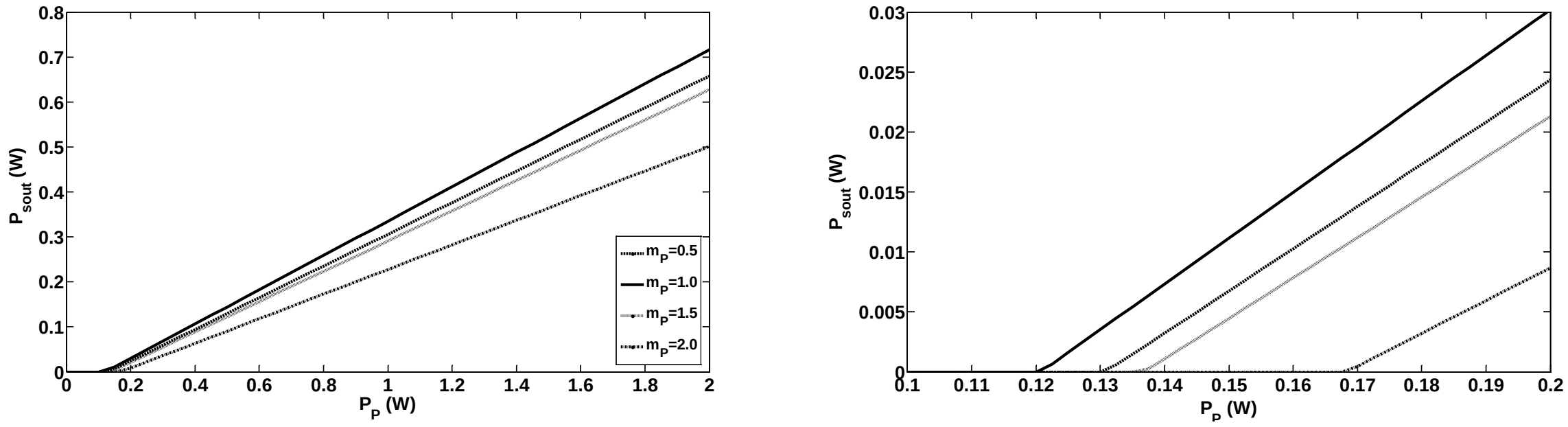

(a) (b)

Fig. 7. (a): Variation of $P_{\text{sout}}$ as functions of $P_{\text{P}}$ with different $m_{\text{P}}$. (b): The values of $P_{\text{sout}}$ that $P_{\text{P}}$ varies from 0.1 to 0.2W.

(6) characteristics of different output reflectivity of fundamental laser $R_{\text{L}}$

Figures 8(a) and 8(b) show the variation of down-converted power $P_{\text{DCout}}$ and efficiency $\eta_{\text{DC}}$ as functions of power of external pump laser $P_{\text{P}}$ with different $R_{\text{L}}$. For simplification, we assume that the idler absorption coefficient $\alpha_{\text{i}}$ in the crystal is 0, the round-trip losses of fundamental laser and signal wave $\beta_{\text{L}}=\beta_{\text{s}}=0$. In the case, the down-converted efficiency $\eta_{\text{DC}}$ reaches 100% for $R_{\text{L}}=0.999$.

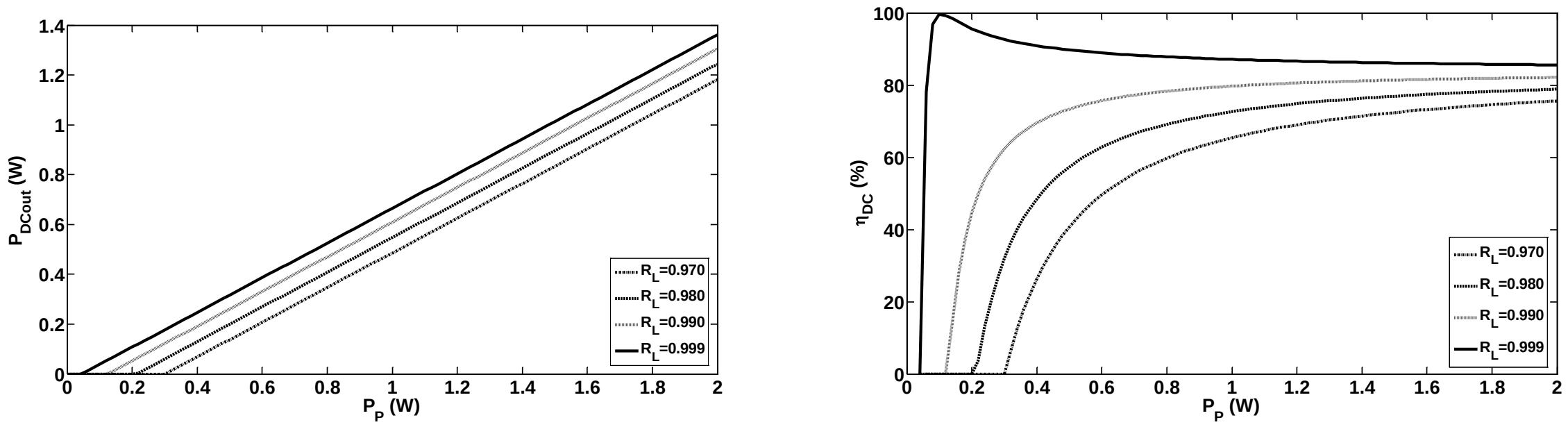

(a) (b)

Fig. 8. (a) and (b): Variation of $P_{\text{DCout}}$ and $\eta_{\text{DC}}$ as functions of $P_{\text{P}}$ with different $R_{\text{L}}$.

(7) characteristics of different output reflectivity of signal wave $R_{\text{s}}$

Figures 9(a) and 9(b) show the variation of output power of signal wave as functions of the power of external pump laser $P_{\text{P}}$ with different $R_{\text{s}}$. In figure (a) $L_{\chi}$=2mm, (b) $L_{\chi}$=6mm. Although smaller $R_{\text{s}}$ leads to higher SPSRO threshold, the slope efficiency of the output signal wave is higher. For certain pump power $P_P$, in order to maximize the output power of signal wave, one must choose a suitable $R_s$.

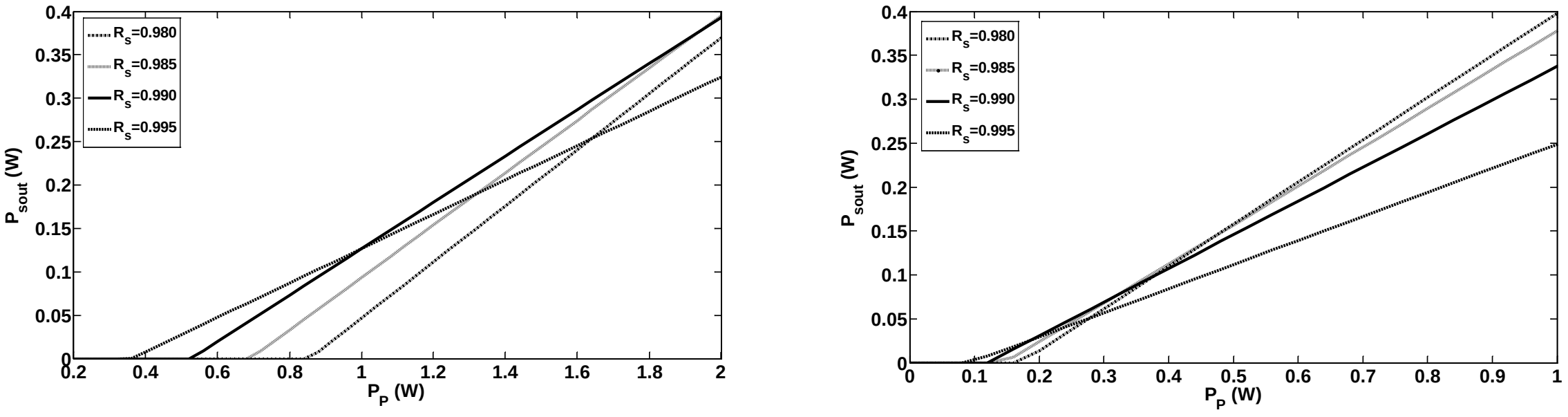

(a) (b)

Fig. 9. Variation of $P_{sout}$ as functions of $P_P$ with different $R_s$. (a) $L_\chi$=2mm; (b) $L_\chi$=6mm.

(8) characteristics of single and double pass pump

Figures 10(a), 10(b), 10(c) and 10(d) show the variation of SPSRO threshold, efficiency of output signal wave $\eta_s$, output power of signal wave $P_{sout}$ and SRO threshold as functions of $R_s$ with single and double pass pump. In this case, the power of external pump laser $P_p$ is 2.0W. As is expected, double pump will increase the efficieny and output power of signal wave, meanwhile, it can reduce the SPSRO threshold. For smaller $R_s$, this positive effects are more obvious than for larger $R_s$.

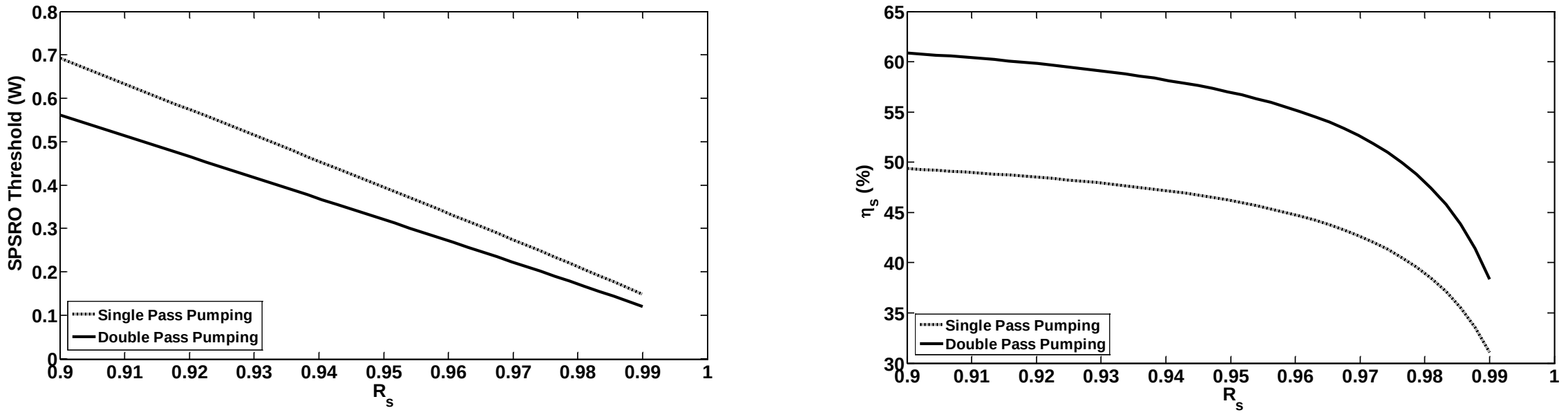

(a) (b)

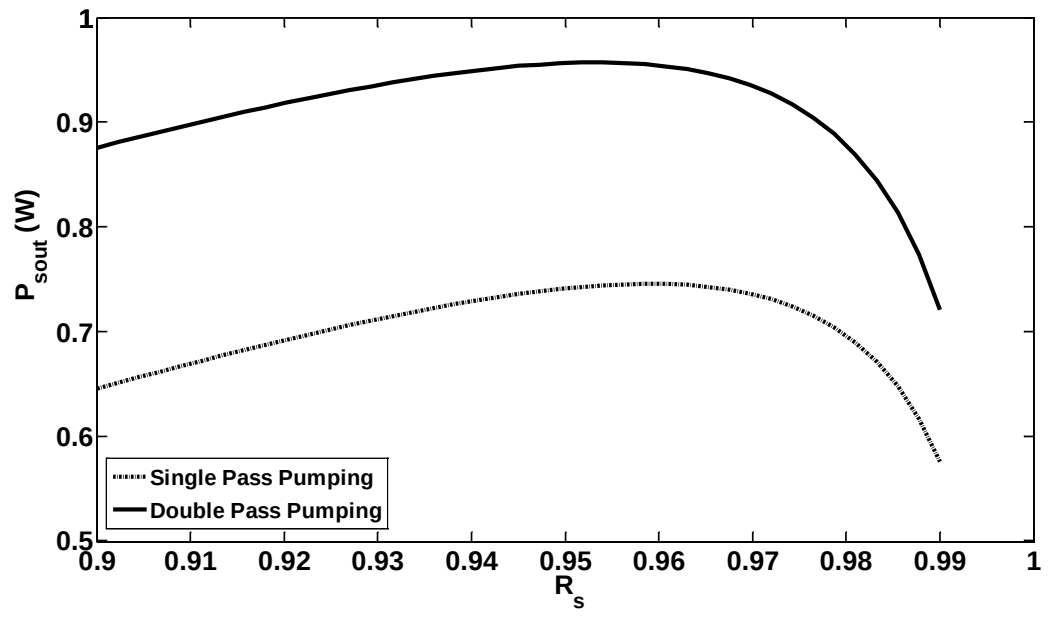

(c)

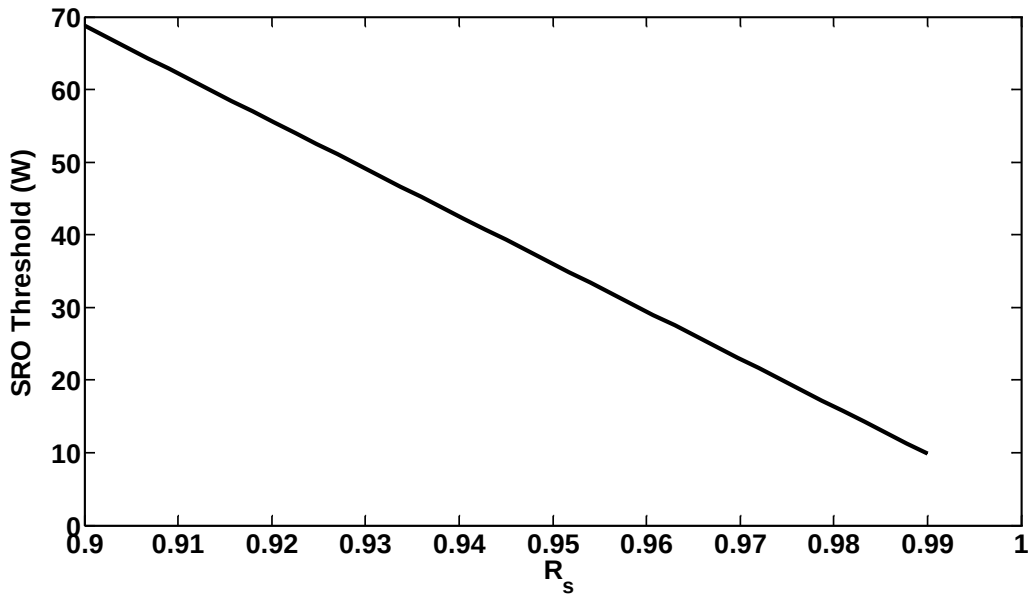

(d)

Fig. 10. (a), (b), (c) and (d): Variation of SPSRO threshold, $\eta_s$, $P_{sout}$ and SRO threshold as functions of $R_s$ with single and double pass pump.

## 4. Conclusion

A model of the steady-state operating, self-pumped singly resonant optical parametric oscillator (SPSRO) has been developed. Employing the $Yb^{3+}$-doped periodically poled lithium niobate co-doped with MgO ($Yb^{3+}$:MgO:PPLN) as examples, we numerically investigate the threshold and efficiency characteristics that the SPSROs operate under different values of the following parameters: phase-mismatching, $Yb^{3+}$ concentration, crystal temperature, $W_{0L}$, $m_P$ , $R_L$, $R_s$, single and double pass pump. For a conventional intracavity SRO, one can tailor the laser gain medium and second-order nonlinear crystal independently. But for a SPSRO based on quasi-three-level laser meduim, the lasing and second order nonlinear processes are generated in the same crystal, and due to the reabsorption process, to reduced the threshold and increase the efficiency, one must choose the crystal length, doped $Yb^{3+}$ concentration and beams' geometry carefully. Lower fundamental laser threshold does not mean lower SPSRO threshold. Meanwhile, higher SPSRO threshold does not mean higher efficiency of the output wave. One must balance between the SPSRO threshold and SRO threshold. The SPSRO characteristics of different parameters must be taken into account for the further optimal analysis.

## Appendix A Parameters in the section two

The parameters of the laser beams are

$$f_L^{+}(x,y,z)=\frac{1}{1+i\tau_L(z)}\exp\left\{-\frac{x^2+y^2}{W_{0L}^{2}[1+i\tau_L(z)]}\right\} \tag{A1}$$

$$|f_P(x,y,z)|^2=\frac{1}{1+[M_P^{2}\tau_P(z)]^2}\exp\left\{-\frac{2(x^2+y^2)}{W_{0P}^{2}\{1+[M_P^{2}\tau_P(z)]^2\}}\right\}\left\{\exp(-\alpha_P z)+R_P\exp[+\alpha_P(z-2L_\chi)]\right\} \tag{A2}$$

$$\tau_L(z)=2(z-f_{0L}L_\chi)/b_L\ ,\ \tau_P(z)=2(z-f_{0P}L_\chi)/b_P \tag{A3}$$

where $x$, $y$, $z$ are the spatial coordinates. $x$=0, $y$=0 are on the axis of the laser crystal, $z$=0 is on the front surface of the crystal. $b_L=k_L W_{0L}^{2}$ is the confocal parameter of the fundamental laser, $W_{0L}$ is the radius of beam waist, $k_L$ is the wave vector. The focusing parameters $\xi_L=L_\chi/b_L$ is used to characterize the focusing. $f_{0L}L_\chi$ is the position of the beam waist of fundamental laser. Similar definitions of the parameters are applied to the external pump laser and signal wave. $M^2$ is the beam quality factor of the external pump laser beam, $\alpha_P$ is the pump absorption coefficient, $R_P$ is the pump reflectivity at the laser output mirror.

$$h_L^{+}(x,y,z)=f_L^{+}(x,y,z)\exp(+ikz),\quad h_L^{-}(x,y,z)=f_L^{+*}(x,y,z)\exp(-ikz) \tag{A4}$$

$$h_L(x,y,z)=h_L^{+}(x,y,z)+h_L^{-}(x,y,z) \tag{A5}$$

$$\eta_\alpha=[1-\exp(-\alpha_P L_\chi)][1+R_P\exp(-\alpha_P L_\chi)] \tag{A6}$$

$$V_L=\frac{\pi}{2}W_{0L}^{2}L_{laser},\quad V_\chi=W_{0L}^{2}L_\chi,\quad V_P=\frac{\pi}{2}W_{0P}^{2},\quad V_{P\alpha}=\frac{\pi}{2}W_{0P}^{2}\frac{\eta_\alpha}{\alpha_P} \tag{A7}$$

$$\Phi_{Lsat}=\frac{n_L V_L}{(f_1^{L}+f_2^{L})c\sigma_L\tau},\quad P_{Lsat}=\hbar\omega_L\frac{c}{2n_L L_{laser}}\Phi_{Lsat} \tag{A8}$$

where $n_L$ and $n_s$ are the refractive indices at the fundamental laser wavelength and signal wavelength, respectively. $\tau$ is the lifetime of the upper laser level, $\sigma_L$ is the emission cross section.

$$F_L(t)=\frac{\pi c\varepsilon_0 n_L|A_L(t)|^2}{4\hbar\omega_L}W_{0L}^{2},\quad \Phi_L(t)=\frac{2F_L(t)L_{laser}}{c/n_L}\ ,\quad P_L=\hbar\omega_L F_L(t) \tag{A9}$$

$$\tau_{Lcavity}=\frac{2L_{laser}n_L/c}{T_L+\beta_L} \tag{A10}$$

where $\beta_L$ is the round-trip loss of the fundamental laser, $T_L$ is the output transmission of the fundamental laser. The definitions of these parameters are also applied to the signal wave.

$$B = -\frac{2\Delta N^0 \sigma L_\chi}{T_L + \beta_L},\ \Delta N^0 = N_2^0 - N_1^0 \tag{A11}$$

$$Q_{LSROth} = \frac{\Phi_{LSROth}}{\Phi_{Lsat}} = \frac{P_{LSROth}}{P_{Lsat}} \tag{A12}$$

$$Q_L = \frac{\Phi_L}{\Phi_{Lsat}} = \frac{P_L}{P_{Lsat}},\quad Q_s = \frac{\Phi_S}{\Phi_{LSROth}} = \frac{P_s}{P_{LSROth}} \frac{\omega_L}{\omega_s} \frac{n_s L_{SRO}}{n_L L_{laser}} \tag{A13}$$

$$P_C = \frac{\hbar\omega_P \cdot V_{P\alpha}}{2(f_1^L + f_2^L)\sigma_L \tau L_{laser}} \tag{A14}$$

The rate $R$ at which ions are excited into the upper laser levels by pump process is

$$R = \eta_P \eta_\alpha \frac{P_P}{\hbar\omega_P} \tag{A15}$$

In the periodically poled quasi-phase-matching (QPM) optical parametric processes, the wave vector mismatch can be written as

$$\Delta k = k_L - k_s - k_i - K_g,\ K_g = \pm 2m\pi / \Lambda_g \tag{A16}$$

where $K_g$ is the grating vector of the $m$th Fourier component, and $\Lambda_g$ is the period of the modulated structure.

The normalized coordinates:

$$p = x / W_{0L},\ q = y / W_{0L},\ s = z / L_\chi,\ s_1 = z_1 / L_\chi,\ s_2 = z_2 / L_\chi \tag{A17}$$

$$m_s = W_{0s} / W_{0L},\ m_P = W_{0P} / W_{0L} \tag{A18}$$

$$d_L(s) = 1 + i\tau_L(z),\ \ d_s(s) = 1 + i\tau_s(z) \tag{A19}$$

$$C_i(s, s_1) = 2i(s - s_1)\xi_L k_L / k_i \tag{A20}$$

$$D_{i1}(s, s_1) = 1/[m_s{}^2 d_s{}^*(s_1)] + 1/d_L(s_1) + 1/C_i(s, s_1) \tag{A21}$$

$$d_{i1}(s, s_1) = d_s{}^*(s_1) d_L(s_1) C_i(s, s_1) D_{i1}(s, s_1) \tag{A22}$$

$$B_{i1}(s,s_1)=\left\{m_s{}^2\cdot d_s{}^*(s_1)+d_L(s_1)\right\}\left\{m_s{}^2\cdot d_s{}^*(s_1)\cdot d_L(s_1)+d_L(s_1)\cdot C_i(s,s_1)+C_i(s,s_1)\cdot m_s{}^2\cdot d_s{}^*(s_1)\right\}^{-1}$$

(A23)

$$D_{L2}(s,s_1)=1/[m_s{}^2 d_s(s)]+1/d_L{}^*(s)+B_{i1}(s,s_1) \tag{A24}$$

$$d_{L2}(s,s_1)=d_{i1}(s,s_1)d_s(s)d_L{}^*(s)D_{L2}(s,s_1) \tag{A25}$$

$$D_{i1}(s_1,s_2)=d_{i1}(1,s_1)d_{i1}{}^*(1,s_2)[B_{i1}(1,s_1)+B_{i1}{}^*(1,s_2)] \tag{A26}$$

$$H_L=2\,\mathrm{Re}\left\{\int_0^1 ds\exp(-i\Delta kLs-\alpha_i Ls/2)\int_0^s ds_1\frac{\exp(i\Delta kLs_1+\alpha_i Ls_1/2)}{d_{L2}(s,s_1)}\right\} \tag{A27}$$

$$H_i=\exp(-\alpha_i L_\chi)\int_0^1 ds_2\int_0^1 ds_1\frac{\exp[i\Delta kL_\chi(s_1-s_2)+\alpha_i L_\chi(s_1+s_2)/2]}{D_{i1}(s_1,s_2)} \tag{A28}$$

where $\alpha_i$ is the absorption coefficient of the idler wave in the crystal.

$$C_L=\frac{4\hbar\omega_L\omega_s\omega_i d^2 L_\chi{}^2}{\varepsilon_0 c n_L{}^2 n_s{}^2 n_i \pi W_{0L}{}^2 L_{SRO} L_{laser} m_s{}^2} \tag{A29}$$

$$C_{LSRO}=\frac{128\pi^2 d^2 L_\chi}{c\varepsilon_0 n_s n_i \lambda_L \lambda_s \lambda_i},\quad C_{iSRO}=\frac{128\pi^2 d^2 L_\chi}{c\varepsilon_0 n_L n_i \lambda_s \lambda_i{}^2} \tag{A30}$$

## References

[1] T. Debuisschert, J. Raffy, J.-P. Pocholle, and M. Papuchon, Intracavity optical parametric oscillator: study of the dynamics in pulsed regime, J. Opt. Soc. Am. B, 13(1996): 1569.

[2] G.A. Turnbull, M.H. Dunn, M. Ebrahimzadeh, Continuous-wave, intracavity optical parametric oscillators: an analysis of power characteristics, Appl. Phys. B 66 (1998), 701.

[3] M. Ebrahimzadeh, G.A. Turnbull, T. J. Edwards, D. J. M. Stothard, I.D. Lindsay, M.H. Dunn, Intracavity continuous-wave singly resonant optical parametric oscillators. J. Opt. Soc. Am. B, 16 (1999) 1499.

[4] A. R. Geiger, H. Hemmati, W. H. Farr, and N. S. Prasad, Diode-pumped optical parametric oscillator, Opt. Lett. 21(1996) 201.

[5] J. Capmany, D. Callejo, V. Bermúdez, E. Diéguez, D. Artigas, L. Torner, Continuous-

wave self-pumped optical parametric oscillator based on $Yb^{3+}$-doped bulk periodically poled $LiNbO_3$(MgO), Appl. Phys. Lett., 79 (2001) 293.

[6] L. Barraco, A. Grisard, E. Lallier, P. Bourdon, and J.-P. Pocholle, Self-optical parametric oscillation in periodically poled neodymium-doped lithium niobate, Opt. Lett. 27(2002) 1540.

[7] M.O. Ramírez, D. Jaque, J. A. Sanz García, L.E. Bausá, J.E. Muñoz Santiuste, 74% Slope efficiency from a diode-pumped $Yb^{3+}$:$LiNbO_3$:MgO laser crystal, Appl. Phys. B 77(2003) 621.

[8] E. Montoya, J. A. Sanz-García, J. Capmany, and L. E. Bausá, A. Diening, T. Kellner and G. Huber, Continuous wave infrared laser action, self-frequency doubling, and tunability of $Yb^{3+}$:MgO:$LiNbO_3$, Appl. Phys. Lett., 87(2000) 4056.

[9] T. Taira, J. Saikawa, T. Kobayashi, R.L. Byer, Diode-pumped tunable Yb:YAG miniature lasers at room temperature modeling and experiment, IEEE J. Selec. Top. Quantum Electron. 3(1997) 100.

[10] W. P. Risk, Modeling of longitudinally pumped solid-state lasers exhibiting reabsorption losses, J. Opt. Soc. Am. B 5 (1988) 1412.

[11] T.Y. Fan, R.L. Byer, Modeling and CW operation of a quasi-three-level 946 nm Nd:YAG laser, IEEE J.Quantum Electron., 23(1987) 605.

[12] S. Guha, Focusing dependence of the efficiency of a singly resonant optical parametric oscillator, Appl. Phys. B 66 (1998), 663.

[13] ChengXian Deng, Zhengjia Li, Changhong Zhu, Focusing dependence of the threshold and efficiency of a singly resonant optical parametric oscillator: generalized analysis, Opt. Commun., 238 (2004) 371.

[14] Hong Shu, Analytic and numeric modeling of diode laser pumped YbYAG laser oscillators and amplifiers. PhD dissertation, University of Central Florida Orlando, Florida, 2003.

[15] E. Montoya, A. Lorenzo, L. E. Bausá, Optical characterization of $LiNbO_3$:$Yb^{3+}$ crystals, J. Phys.: Condens. Matter 11 (1999) 311.